\documentclass{gGAF2e}

\renewcommand{\Omega}{{\varOmega}}

\begin{document}

\jvol{00} \jnum{00} \jyear{2018} 

\markboth{\rm O.~GOEPFERT AND A.~TILGNER}{\rm GEOPHYSICAL AND ASTROPHYSICAL FLUID DYNAMICS}


\title{Mechanisms for magnetic Field generation in precessing cubes}

\author{O. GOEPFERT and A. TILGNER $^{\ast}$\thanks{$^\ast$Corresponding author. Email: Andreas.Tilgner@phys.uni-goettingen.de
\vspace{6pt}}\\
\vspace{6pt}  
Institute of Geophysics, University of G\"ottingen, 37077 G\"ottingen,
Germany\\
\vspace{6pt}\received{2017} }

\maketitle

\begin{abstract}
It is shown that flows in precessing cubes develop at certain parameters large
axisymmetric components in the velocity field which are large enough to either generate
magnetic fields by themselves, or to contribute to the dynamo effect if inertial modes are
already excited and acting as a dynamo. This effect disappears at small Ekman
numbers. The critical magnetic Reynolds number also increases at low Ekman
numbers because of turbulence and small scale structures.

\begin{keywords} Precession; Kinematic dynamos
\end{keywords}

\end{abstract}

\section{Introduction}

Precession driven flow is known to lead to magnetic field generation since the
first dynamo simulations in precessing spheres \citep{Tilgne05}. The mechanisms
exciting flows suitable for dynamo action identified in this work were Ekman
pumping at the boundaries and triad resonances. The mechanism based on Ekman
pumping is superseded at small Ekman numbers by the triad resonances which result
from the coupling of inertial waves and which are a bulk instability. Dynamos
based on this instability are expected in containers of any shape. They were for
instance observed in precessing cubes \citep{Goepfe16}.

A laboratory experiment is currently under construction which intends to
demonstrate the dynamo effect in precessing sodium filled vessels
\citep{Stefan15}. The construction of the experiment allows for different vessel
geometries, but a cylindrical container is the simplest choice and will be
realized first. Dynamos in cylinders are little explored as of now.
\citet{Nore11} found dynamos in laminar flows within precessing cylinders. More
recently, \citet{Giesec18} obtained in the same system dynamos based on
axisymmetric flows which resemble very much the kinematic dynamos introduced by
\citet{Dudley89} and which inspired the VKS experiment \citep{Moncha07}. This
discovery motivates us to revisit the problem of the precessing cube and to
search for an additional dynamo mechanism in this geometry. The choice of
parameters and the issues addressed in the present paper are clearly guided by
the laboratory application and not by some astrophysical object.

Section \ref{Model} formulates the problem to be simulated. Section
\ref{Hydrodynamics} describes hydrodynamic processes and the observation of
flows analogous to the axisymmetric flows in \citet{Giesec18}. Section
\ref{Kinematic} finally deals with dynamo action within these flows.

\section{The mathematical model of a precessing cube}
\label{Model}

A cube of side length $L$ filled with incompressible liquid of density
$\rho$ and viscosity $\nu$ rotates with angular frequency $\tilde \omega_D$
about the $x$-axis and precesses with angular frequency $\tilde \Omega_P$. The
$x$-axis is part of a Cartesian tripod attached to the cube, whose sides are
parallel to $x$, $y$ and $z$-axes. The index $D$ in $\tilde \omega_D$ stands for
diurnal rotation, a term borrowed from the geophysical application
\citep{Tilgne15}. The precession axis forms the angle $\alpha$ with the $x$-axis.

There are several reasonable options for removing dimensions from the governing
equations. Here, we adopt the choice already made in \citet{Goepfe16} and base
the unit of time on the total angular frequency of rotation about the container
axis, denoted as $x$-axis, to which both $\tilde \omega_D$ and $\tilde
\Omega_P$ contribute. The unit of time is then
$1/(\tilde \omega_D + \tilde \Omega_P \cos \alpha)$
and the nondimensional rotation rates $\omega_D$ and $\Omega_P$ derived form
$\tilde \omega_D$ and $\tilde \Omega_P$ are
\begin{align}
\omega_D\, =\,&\, \frac{\tilde \omega_D}{\tilde \omega_D + \tilde \Omega_P \cos \alpha}\,
= \,\frac{1}{1+\Omega \cos \alpha}\,,\\
\Omega_P\, = \,&\, \frac{\tilde \Omega_P}{\tilde \omega_D + \tilde \Omega_P \cos \alpha}\,
=\,\frac{\Omega}{1+\Omega \cos \alpha}
\end{align}
with $\Omega=\Omega_P/\omega_D={\tilde \Omega_P}/{\tilde \omega_D}$.
Let hats denote unit vectors. In the $x,y,z-$frame, which we will call the
``boundary frame'' from now on, the vector of precession $\bm \Omega_P$ is given
by
\begin{equation}
\bm \Omega_P\, =\, \Omega_P \cos \alpha ~ \hat{\bm x} + \Omega_P ~ \bm p(t)
\end{equation}
with
\begin{equation}
\bm p(t)\,=\,\sin \alpha (\cos \omega_D t  ~ \hat{\bm y}-\sin \omega_D t ~  \hat{\bm z})\,.
\end{equation}
The equation of motion for the (non-dimensional) velocity $\bm v(\bm r,t)$ as a
function of position $\bm r$ and time $t$ and the pressure $\phi (\bm r,t)$ reads in
the frame attached to the cube
\begin{align}
  \upartial_t \bm v +(\bm v {\bm \cdot} \bm \nabla) \bm v + 2 (\hat{\bm x}+\Omega_P \bm p(t)) \times \bm v = \,
  &\, - \bm \nabla \phi+ \mathrm{Ek} \nabla^2 \bm v + \Omega_P \omega_D (\hat{\bm x} \times \bm p(t)) \times \bm r\,,
\hskip 8mm \label{eq:NS_BQ}\\
%
%
\bm \nabla {\bm \cdot} \bm v\, =\, &\,0
\label{eq:conti_BQ}
\end{align}
with an Ekman number $\mathrm{Ek}$ given by
\begin{equation}
\mathrm{Ek}\,=\,\frac{\nu}{(\tilde \omega_D + \tilde \Omega_P \cos \alpha) L^2}\,.
\end{equation}

It proved useful already in \citet{Goepfe16} to use a finite difference code
implemented on GPUs to simulate precession driven flow in cubes. In order to
take full advantage of the special architecture of GPUs, this method avoids the
need for any Poisson solver by simulating the flow of a weakly compressible
fluid \citep{Tilgne12b}. If $c$ is the sound speed, this method replaces $\bm \nabla {\bm \cdot} \bm v=0$
with the linearized continuity
equation $\upartial_t \rho + \bm \nabla {\bm \cdot} \bm v = 0$ and the term $-\bm \nabla \phi$ in
(\ref{eq:NS_BQ}) becomes $-c^2 \bm \nabla \rho$. The equations actually solved
by
the finite difference scheme are
\begin{align}
\upartial_t \bm v +(\bm v {\bm \cdot} \bm \nabla) \bm v + 2 (\hat{\bm x} + \Omega_P \bm p(t)) \times \bm v\, = \,&\,
- c^2 \bm \nabla \rho+ \mathrm{Ek} \nabla^2 \bm v + \Omega_P \omega_D (\hat{\bm x} \times \bm p(t)) \times \bm r\,,
\hskip 8mm \label{eq:NS}\\
%
%
\upartial_t \rho + \bm \nabla {\bm \cdot} \bm v\, =\,&\, 0\,.
\label{eq:conti}
\end{align}

The sound speed $c$ is chosen to keep the Mach number $|\bm v|/c$ below $0.04$
everywhere. In addition, $c$ needs to be large enough so that the time it takes
sound waves to travel across the cube is much less than the rotation period,
which expressed in the non-dimensional quantities requires $c \gg 2\pi$. In the
simulations presented here, $c^2 = 500$. The simulations are started from
$\rho=1$ and $|\rho-1|$ stays below $5 \times 10^{-4}$ during the course of the
computations for this choice of $c^2$.
The finite value of $c^2$ should then have insignificant effects
for the purposes of this paper. To confirm this, 
$c^2$ was varied form 500 to 5000 for $\mathrm{Ek}=10^{-4}$ and $\Omega_P=-0.018$,
and the kinetic energy density $E_{\mathrm{kin}}$, to be defined below, 
was $1.704 \times 10^{-2} \pm 3 \times 10^{-5}$ in all cases.

For the kinematic dynamo problem, the
induction equation for the magnetic field $\bm B (\bm r,t)$
\begin{equation}
\upartial_t \bm B +\bm \nabla \times (\bm B \times \bm v)\, =\,
\frac{\mathrm{Ek}}{\mathrm{Pm}} \nabla^2 \bm B   \,,\hskip 15mm
\bm \nabla {\bm \cdot} \bm B \,=\, 0
\label{eq:induc_BQ}
\end{equation}
is solved together with the equations of motion,
where the magnetic Prandtl number $\mathrm{Pm}$ is given by
$\mathrm{Pm}={\nu}/{\lambda}$
with $\lambda$ the magnetic diffusivity of the fluid.

Free slip conditions are applied to the velocity field at the boundaries. 
These enforce that the velocity component normal to a boundary and the normal
derivative of the tangential components vanish on the boundary. As in
other studies of dynamos in non spherical geometry, we use
boundary conditions for the magnetic field which can be expressed locally
\citep{Krauze10,Cebron12b,Giesec15,Giesec18}, namely the
pseudo-vacuum boundary conditions which require the tangential
components of $\bm B$ to be zero at the boundaries.

The investigated parameter range is essentially the same as in \citet{Goepfe16}
(see table \ref{table1})
with some points added at large $\mathrm{Ek}$ ($\mathrm{Ek}=10^{-3}$) and large
$|\Omega|$. However, the computations were not extended to
computationally more demanding parameters than previously, in particular not to
small $\mathrm{Ek}$.

\begin{table}\centering
\begin{tabular}{ c | c | c | c }

$\mathrm{Ek}$ & $N$ & $\Omega_P$  &  $\mathrm{Pm}$ \\
\hline
 
$10^{-3}$    &    $64$    &    -0.02\dots-0.3    &    1\dots50    \\
\hline

$2.5\cdot10^{-4}\dots10^{-4}$    &    $128$    &    -0.02 \dots -0.3    &  
0.1\dots30    \\
\hline

$7.5\cdot10^{-5}\dots10^{-5}$    &    $256$    &    -0.02\dots-0.1, -0.16,
-0.3    &   0.1\dots10    \\

\end{tabular}
\caption{Overview of the parameters used in the simulations. $N$ is the number
of grid points in each Cartesian direction. The angle $\alpha$ is always set to
$\alpha=60^o$. Detailed parameters can be deduced from the figures.}

\label{table1}
\end{table}

\section{Hydrodynamics}
\label{Hydrodynamics}

\begin{figure}
\begin{center}
\begin{minipage}{100mm}
\subfigure[$s_1t_1$]{
\resizebox*{5cm}{!}{\includegraphics{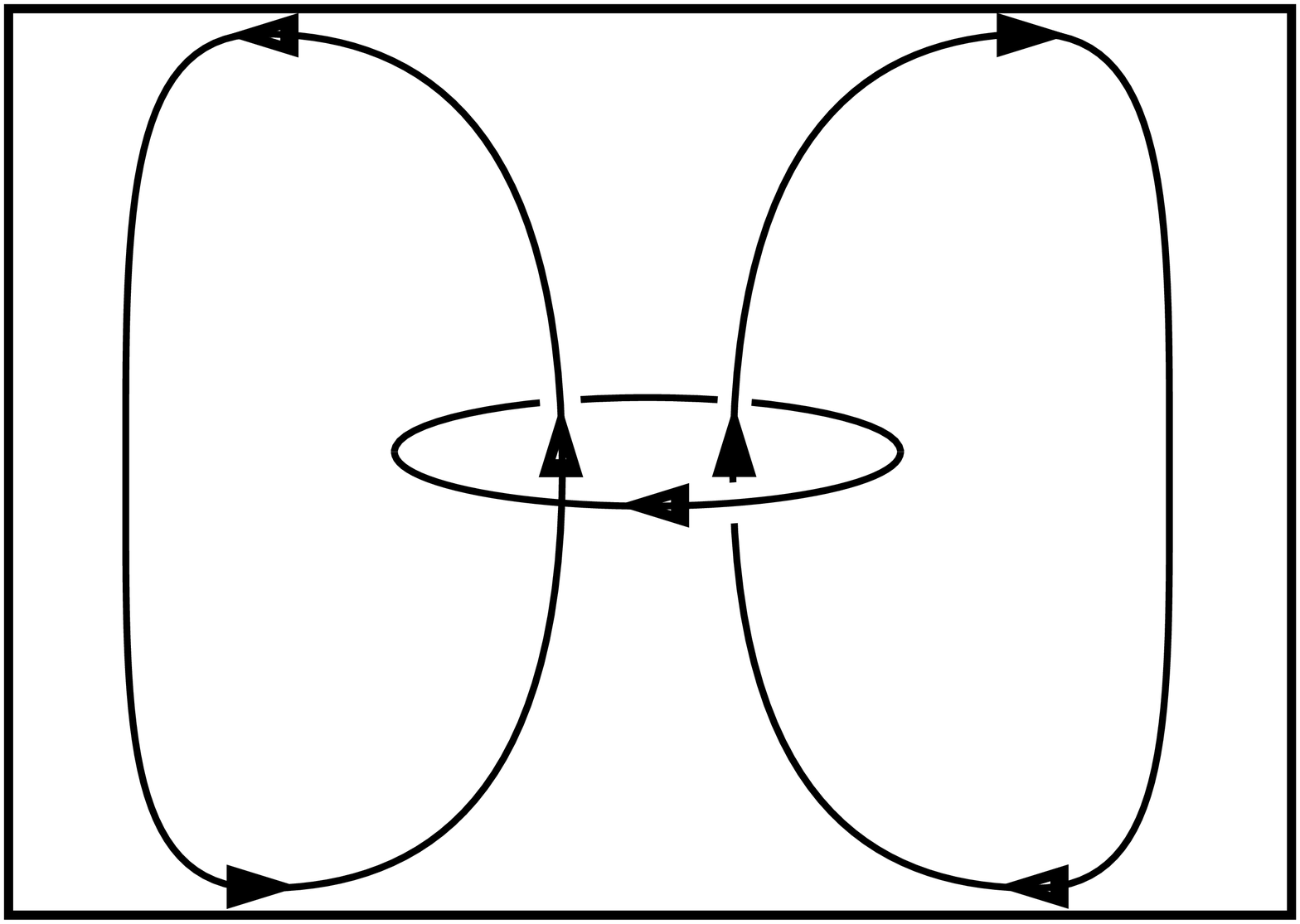}}}%
\subfigure[$s_2t_1$]{
\resizebox*{5cm}{!}{\includegraphics{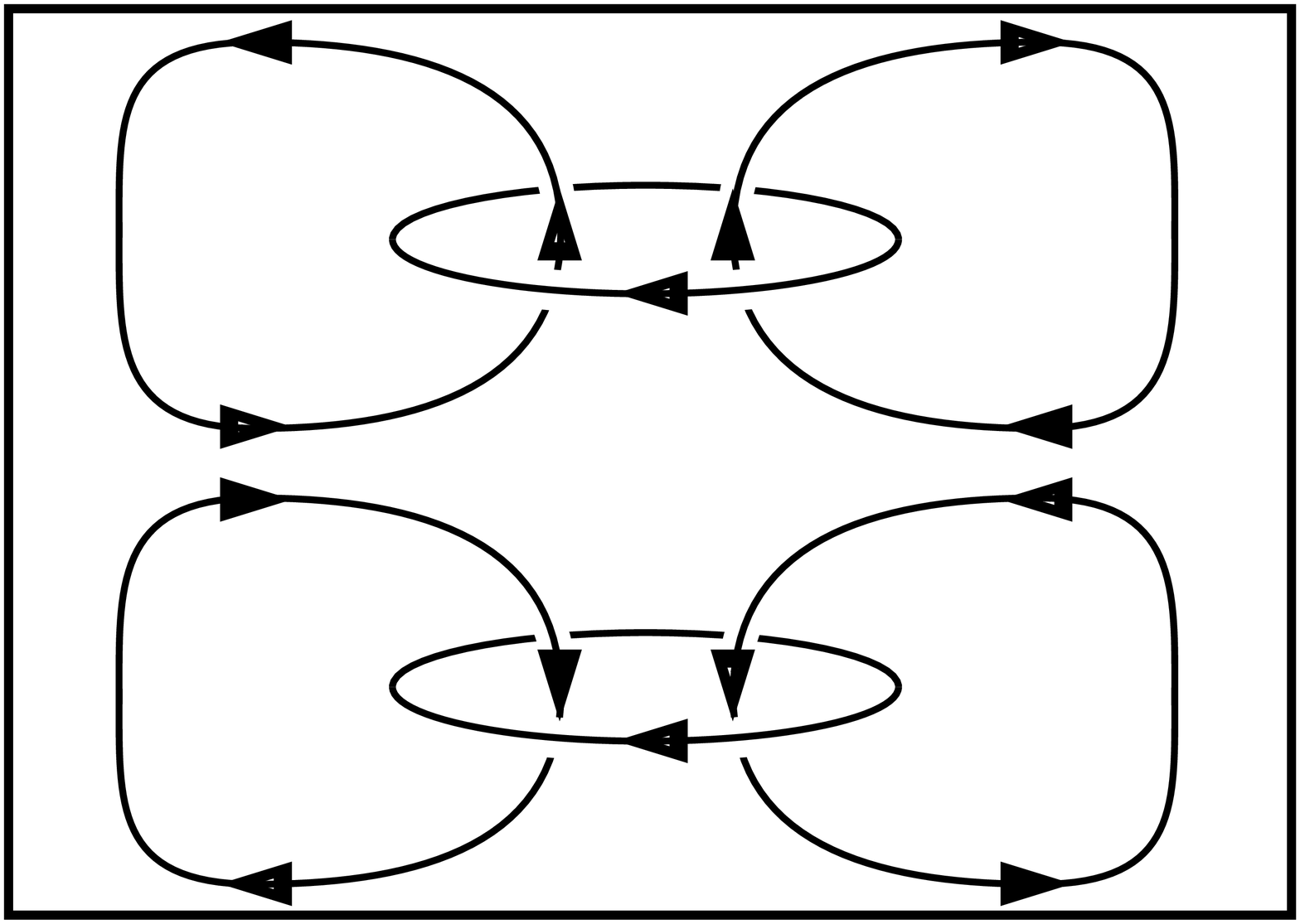}}}%
\caption{Sketch of the $s_1t_1$ (a) and $s_2t_1$ (b) flows.}%
\label{fig1}
\end{minipage}
\end{center}
\end{figure}

The main purpose of this section is to show that precessional flow in cubes can
contain large fractions of nearly axisymmetric flow, exact axisymmetry being
impossible because of the corners and edges of the cube. These axisymmetric
flows furthermore have the same topology as some flows studied for their
kinematic dynamo properties by \citet{Dudley89}. These flows consist of a
rotation about a central axis and a meridional circulation built from either one
or two tori, designated as $s_1t_1$ and $s_2t_1$ flow, respectively, in keeping
with the notation introduced in \citet{Dudley89}. These flows are sketched in
figure~\ref{fig1}.

In order to construct objective diagnostics for the presence of these flows,
consider the following definitions: The energy density of the flow, 
$E_{\mathrm{kin}}$, is defined as 
$E_{\mathrm{kin}}={V}^{-1} \big\langle \int \frac{1}{2} \bm v^2 {\mathrm d}V \big\rangle$
where $\langle \cdots \rangle$ denotes average over time and the integration
extends over the entire fluid volume $V$. Note that $V=1$ in our geometry. It
will also be useful to consider the part of the velocity field $\bm v_a$ which
is antisymmetric with respect to reflection at the origin,
\begin{equation}
\bm v_a\,=\,\tfrac{1}{2}\bigl(\bm v(\bm r)+ \bm v(-\bm r)\bigr)
\end{equation}
and its energy
%
\begin{equation}
E_a\,=\,\frac{1}{V} \biggl\langle\, \int \tfrac{1}{2} \bm v_a^2\, {\mathrm d}V \biggr\rangle\,.
\label{eq:E_a}
\end{equation}
These quantities were used in the past for detecting instability. In
the present context, they are also of interest because $E_a=0$ for the $s_2t_1$
flow, whereas $E_a \neq 0$ for the $s_1t_1$ flow because of its meridional components.
It is however more intuitive to distinguish the $s_1t_1$ and $s_2t_1$ flows thanks to
a mirror symmetry. Let us define the velocity field $\bm v_e$ which is
the part of $\bm v$ which is mirror symmetric with respect to the plane 
perpendicular to the rotation axis $x$ and which divides the cube in two equal halves:
\begin{align*}
v_{es}(s,\varphi,x)\, =\,&\,
\tfrac{1}{2} \bigl( v_s(s,\varphi,x) + v_s(s,\varphi,-x) \bigr)\,,\\
%
v_{e\varphi}(s,\varphi,x)\, =\,&\,
\tfrac{1}{2} \bigl( v_\varphi(s,\varphi,x) + v_\varphi(s,\varphi,-x) \bigr)\,,\\
%
v_{ex}(s,\varphi,x)\, =\,&\,
\tfrac{1}{2} \bigl( v_x(s,\varphi,x) - v_x(s,\varphi,-x) \bigr)\,.
\end{align*}
where $(s,\varphi,x)$ are cylindrical coordinates with the $x$-axis as
distinguished axis.
The index $e$ stands for equatorially symmetric because of the obvious analogy
with equatorially symmetric flows in spheres.
$\bm v_e = \bm v$ for the $s_2t_1$ flow, whereas the $s_1t_1$ flow has again
mixed symmetry.

We next have to separate the axisymmetric components from the others. We
obtain the axisymmetric contributions to the velocity
components $v_s$, $v_\varphi$, $v_x$ from the integral
\begin{equation}
v_{x0}(s,x) \,= \, \frac{1}{2\pi} \int_0^{2\pi} v_x(s,\varphi,x)  \,{\mathrm d}\varphi
\bigg/ \left( \frac{1}{2\pi} \int_0^{2\pi} I(s,\varphi,x)  \,{\mathrm d}\varphi \right)
\label{phi-average}
\end{equation}
and likewise for $v_{s0}$, $v_{\varphi 0}$ and the axisymmetric and equatorially
symmetric components
$v_{ex0}$, $v_{es0}$, $v_{e\varphi 0}$. The arguments of all these quantities
have to span the intervals $0 \leq x \leq 1$ and $0 \leq s \leq \sqrt{2}/2$. The
integration in (\ref{phi-average}) extends over regions partly outside the
cube for $1/2 < s \leq \sqrt{2}/2$. The average in (\ref{phi-average}) is
intended to be an average over the cube, so that $v_x$ is set to zero outside
the cube, and the function $I(s,\varphi,x)$ is 1 within the cube and zero
outside. The azimuthally averaged velocities are finally transformed into
energies as for example in
\begin{equation}
E_{x0} \,= \,\frac{1}{V} \biggl\langle\, \int \tfrac{1}{2} v_{x0}^2\,{\mathrm d} V \biggr\rangle
\end{equation}
and similarly for $E_{s0}$, $E_{\varphi 0}$ and $E_{ex0}$, $E_{es0}$,
$E_{e\varphi 0}$.

\begin{figure}
\begin{center}
\begin{minipage}{160mm}
\subfigure[$$]{
\resizebox*{8cm}{!}{\includegraphics{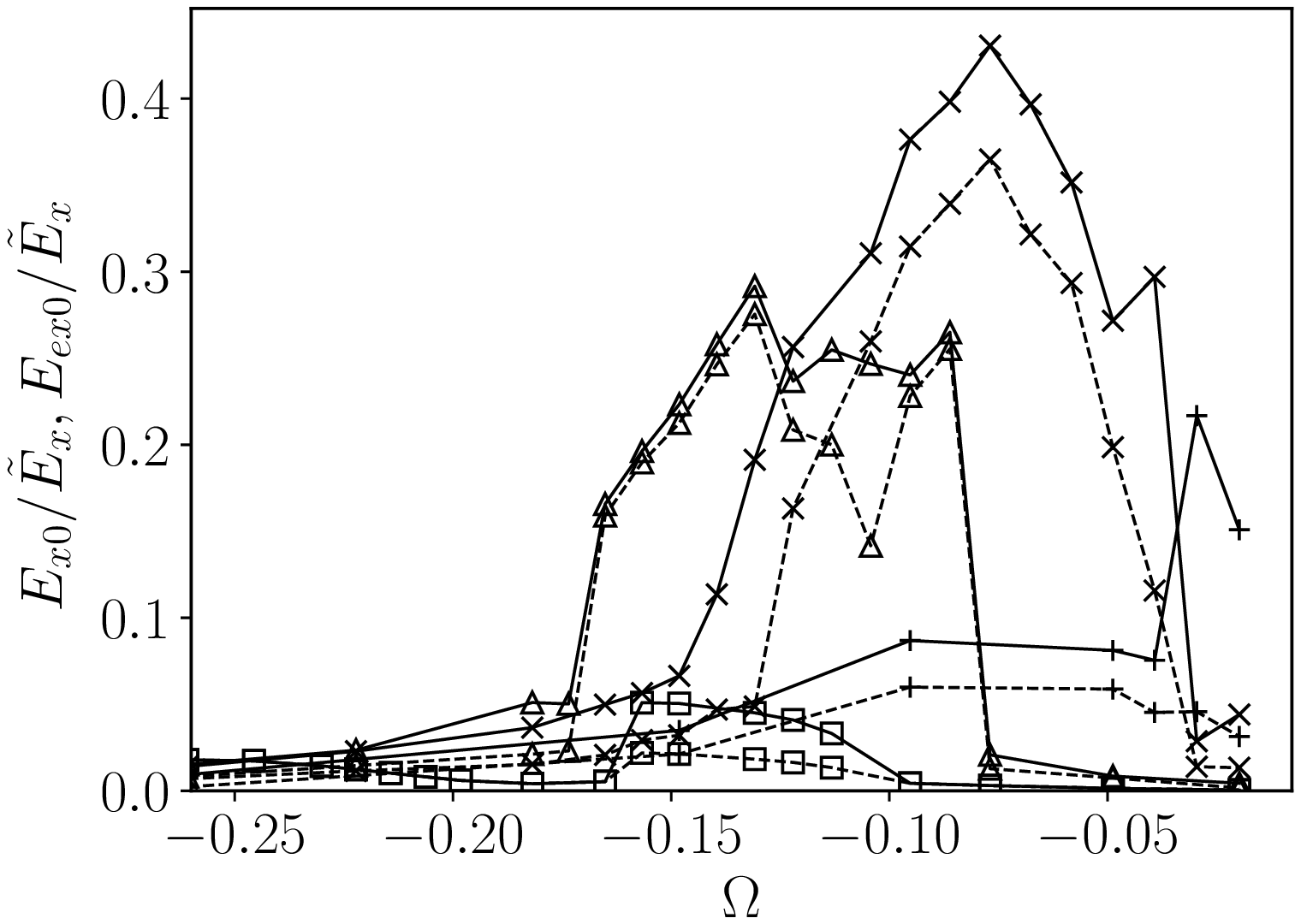}}}%
\subfigure[$$]{
\resizebox*{8cm}{!}{\includegraphics{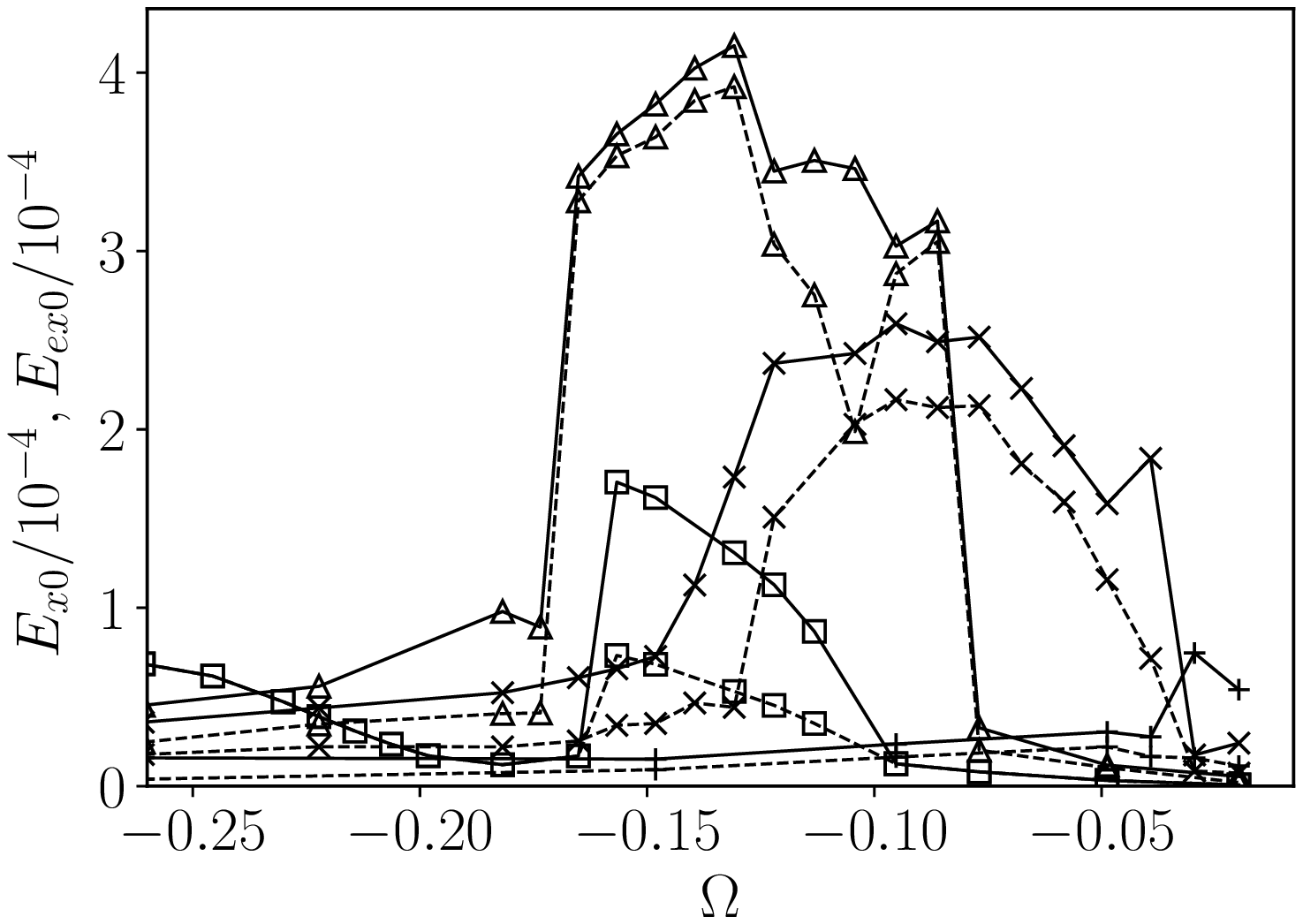}}}%
\caption{(a) $E_{x0}/\tilde E_x$ as a function of $\Omega$ for
$\mathrm{Ek}=1.0\cdot10^{-3}$ ($\square$),$\mathrm{Ek}=2.5\cdot10^{-4}$ ($\triangle$),
$\mathrm{Ek}=1.0\cdot10^{-4}$ ($\times$) and $\mathrm{Ek}=1.0\cdot10^{-5}$ ($+$).
The continuous lines connect points for $E_{x0}/\tilde E_x$, and the dashed lines are
for $E_{ex0}/\tilde E_x$.
(b) $E_{x0}$ as a function of $\Omega$ with the same symbols as in (a).}%
\label{fig2}
\end{minipage}
\end{center}
\end{figure}

\begin{figure}
\begin{center}
\begin{minipage}{160mm}
\subfigure[$$]{
\resizebox*{8cm}{!}{\includegraphics[width=100mm]{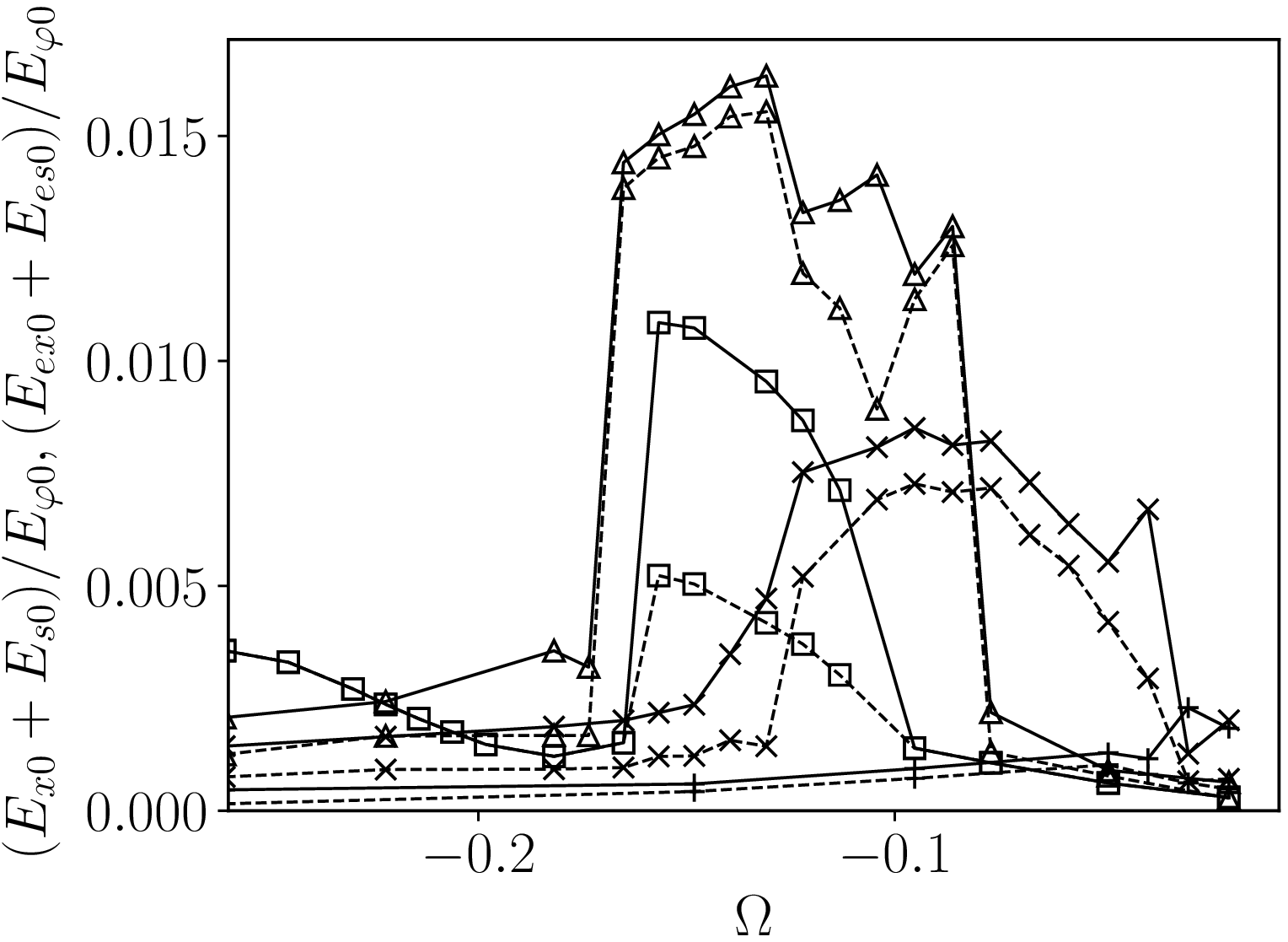}}}%
\subfigure[$$]{
\resizebox*{8cm}{!}{\includegraphics{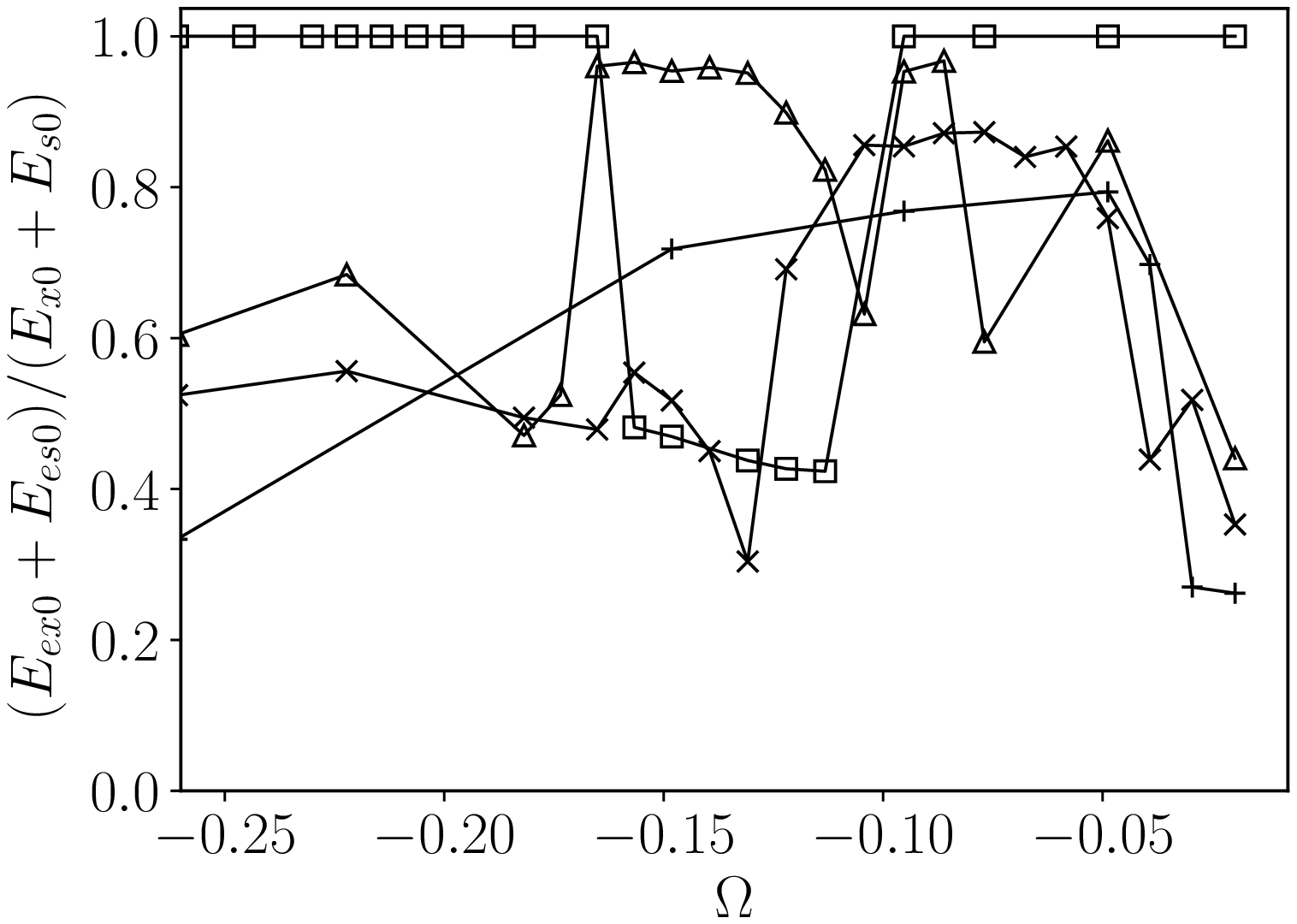}}}%
\caption{(a) $(E_{x0}+E_{s0})/E_{\varphi 0}$ and
$(E_{ex0}+E_{es0})/E_{\varphi 0}$ as a function of $\Omega$ for
$\mathrm{Ek}=1.0\cdot10^{-3}$ ($\square$),$\mathrm{Ek}=2.5\cdot10^{-4}$ ($\triangle$),
$\mathrm{Ek}=1.0\cdot10^{-4}$ ($\times$) and $\mathrm{Ek}=1.0\cdot10^{-5}$ ($+$).
The continuous lines connect points for $(E_{x0}+E_{s0})/E_{\varphi 0}$, and the
dashed lines are for $(E_{ex0}+E_{es0})/E_{\varphi 0}$.
(b) $(E_{ex0}+E_{es0})/(E_{x0}+E_{s0})$ as a function of $\Omega$ with the same
symbols as in (a).}
\label{fig3}
\end{minipage}
\end{center}
\end{figure}

Yet another quantity appears in figure~\ref{fig2}, which is $\tilde E_x$, the energy
contained in the non-axisymmetric components of $v_x$:
\begin{equation}
\tilde E_x\, =\, \frac{1}{V} \biggl\langle\, \int \tfrac{1}{2} (v_x-v_{x0})^2 \,{\mathrm d}V \biggr\rangle\,.
\end{equation}

Figure~\ref{fig2}(a) plots $E_{x0}/\tilde E_x$ as a function of $\Omega$ for different
$\mathrm{Ek}$. This quantity detects a dramatic increase of the axisymmetric
components at some $\Omega$. This increase is not spread equally among the 
velocity components, as revealed by figure~\ref{fig3}(a). This figure shows
$(E_{x0}+E_{s0})/E_{\varphi 0}$ as a function of $\Omega$ and thus compares
axisymmetric meridional and azimuthal components. Large values in
figures~\ref{fig2} and \ref{fig3}(a) correlate with each other, which means that if a large
axisymmetric component appears at some $\Omega$, it appears because the
axisymmetric meridional components have increased.

Precession forces a basic flow in the container frame which is approximately a solid
body rotation about an axis other than the rotation axis of the container. This
flow thus contains already through direct forcing and without intervening instability
non axisymmetric components which contribute to $\tilde E_x$, and axisymmetric
components which contribute to $E_{\varphi 0}$. Figures \ref{fig2}(a) and
\ref{fig3}(a) show broadly the same variation because $\tilde E_x$ and $E_{\varphi
0}$ both are dominated by the basic flow which exists at all $\Omega$, whereas $E_{s0}$ and
$E_{x0}$ have significant magnitude only in certain intervals of $\Omega$.
For comparison, figure~\ref{fig2}(b) shows $E_{x0}$ without normalization
with $\tilde E_x$ exhibits rapid variations as a function of
$\Omega$ at the same $\Omega$ as figure \ref{fig2}(a).

Figure~\ref{fig3}(a) also shows $(E_{ex0}+E_{es0})/E_{\varphi 0}$. This ratio exactly
coincides with $(E_{x0}+E_{s0})/E_{\varphi 0}$ if the axisymmetric part of the
flow is purely of the $s_2t_1$ type. If the two ratios differ, there is a contribution
to the axisymmetric flow by the opposite symmetry, whose simplest
representative is the $s_1t_1$ flow. There is generally some admixture of both
symmetries. For a quantitative measure, figure~\ref{fig3}(b) plots
$(E_{ex0}+E_{es0})/(E_{x0}+E_{s0})$. This ratio is 1 in an $s_2t_1$ flow and 0 in a 
pure $s_1t_1$ flow. Figure~\ref{fig3}(b) shows that the $s_2t_1$ flow clearly
dominates the axisymmetric flow at some parameters, while it contributes less than one
half of the axisymmetric meridional flow at other parameters.

\begin{figure}
\begin{center}
\begin{minipage}{100mm}
\subfigure[$$]{
\resizebox*{5cm}{!}{\includegraphics{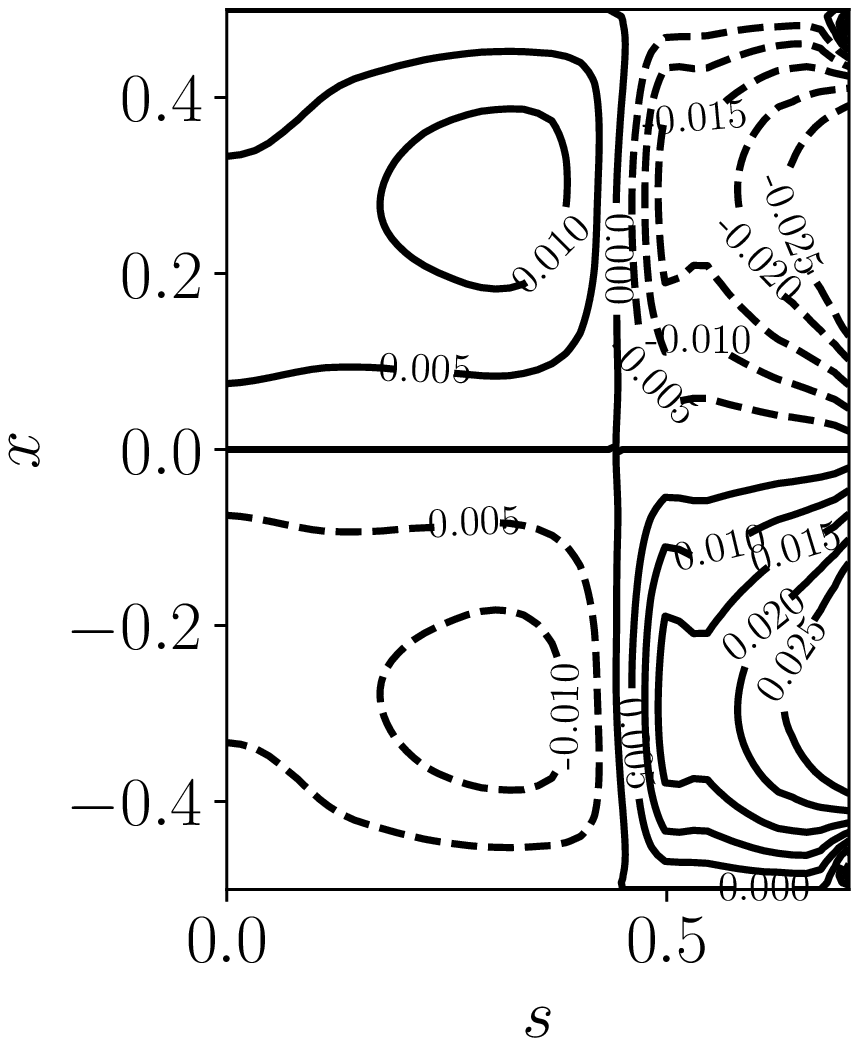}}}%
\subfigure[$$]{
\resizebox*{5cm}{!}{\includegraphics{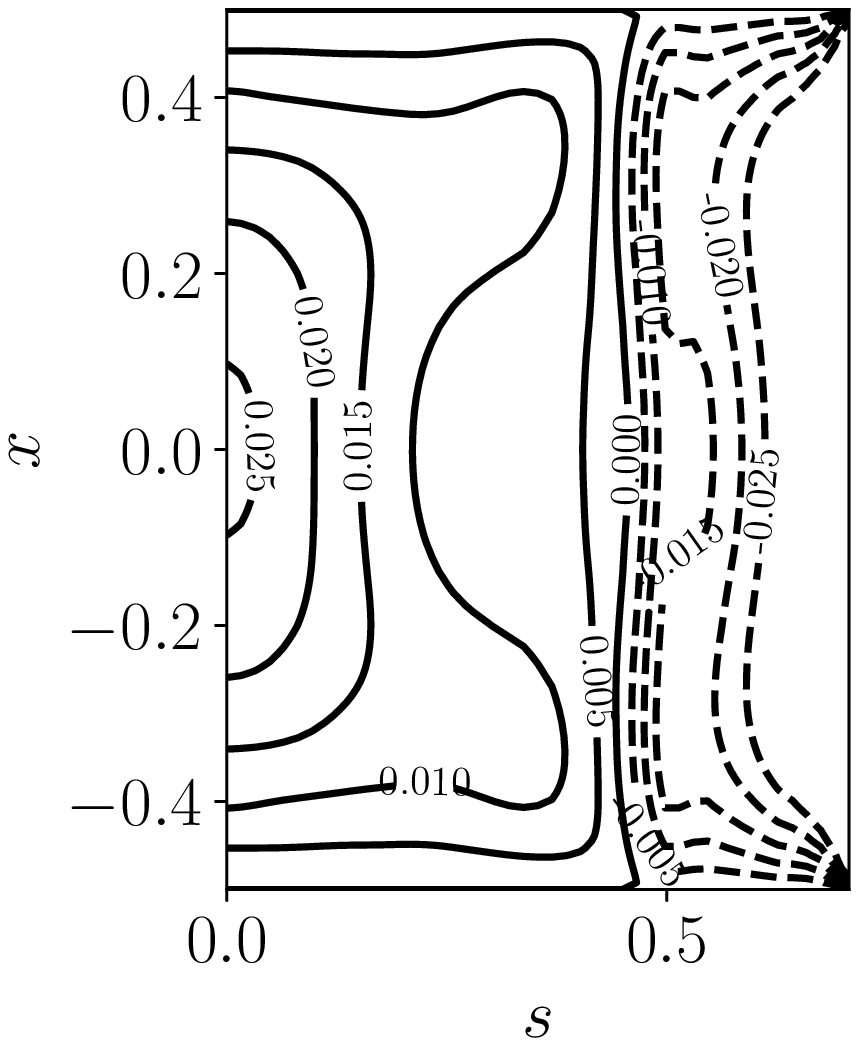}}}%
\caption{Contour plots of (a) $v_{ex0}$ and (b) $v_{x0}-v_{ex0}$, 
at $\mathrm{Ek}=10^{-4}$ and $\Omega=-0.04$}%
\label{fig4}
\end{minipage}
\end{center}
\end{figure}

Visualizations such as in figure~\ref{fig4}
confirm that the sketches in
figure~\ref{fig1} qualitatively represent the actual flows. Figure~\ref{fig4} shows
contour plots of the equatorially antisymmetric and symmetric part of $v_{x0}$ at $\mathrm{Ek}=10^{-4}$ and $\Omega=-0.04$. 
One recognizes the $s_2t_1$ and $s_1t_1$ patterns but one
also notices that $v_{x0}$ has a single sign for $0 \leq s \lesssim 1/2$ in each
half of the cube, so that the return flows must mostly occur near the edges of
the cube in the region $1/2 \leq s \leq \sqrt{2}/2$.

\begin{figure}
\begin{center}
\begin{minipage}{140mm}
\subfigure{
\resizebox*{7cm}{!}{\includegraphics{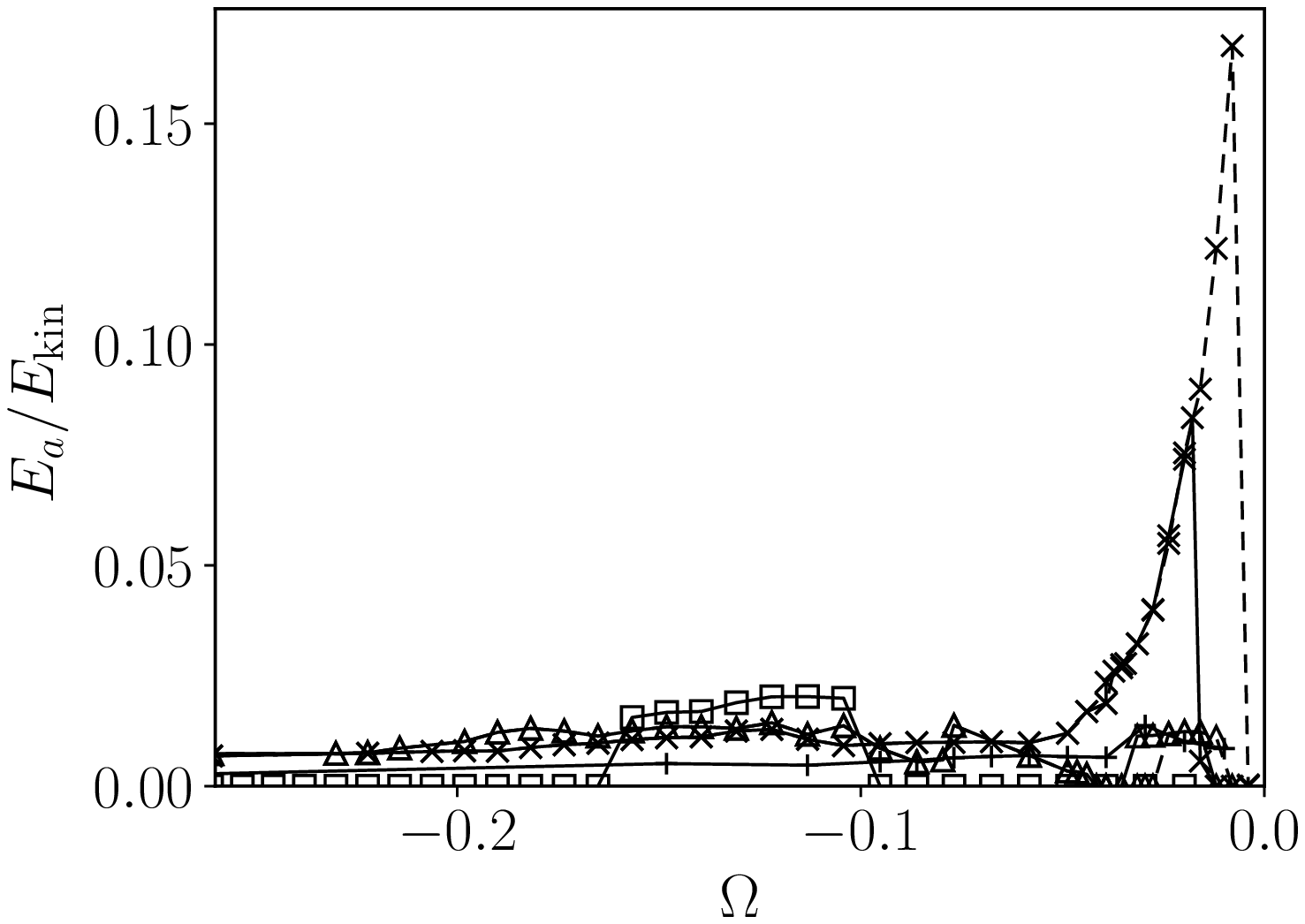}}}%
\subfigure{
\resizebox*{7cm}{!}{\includegraphics{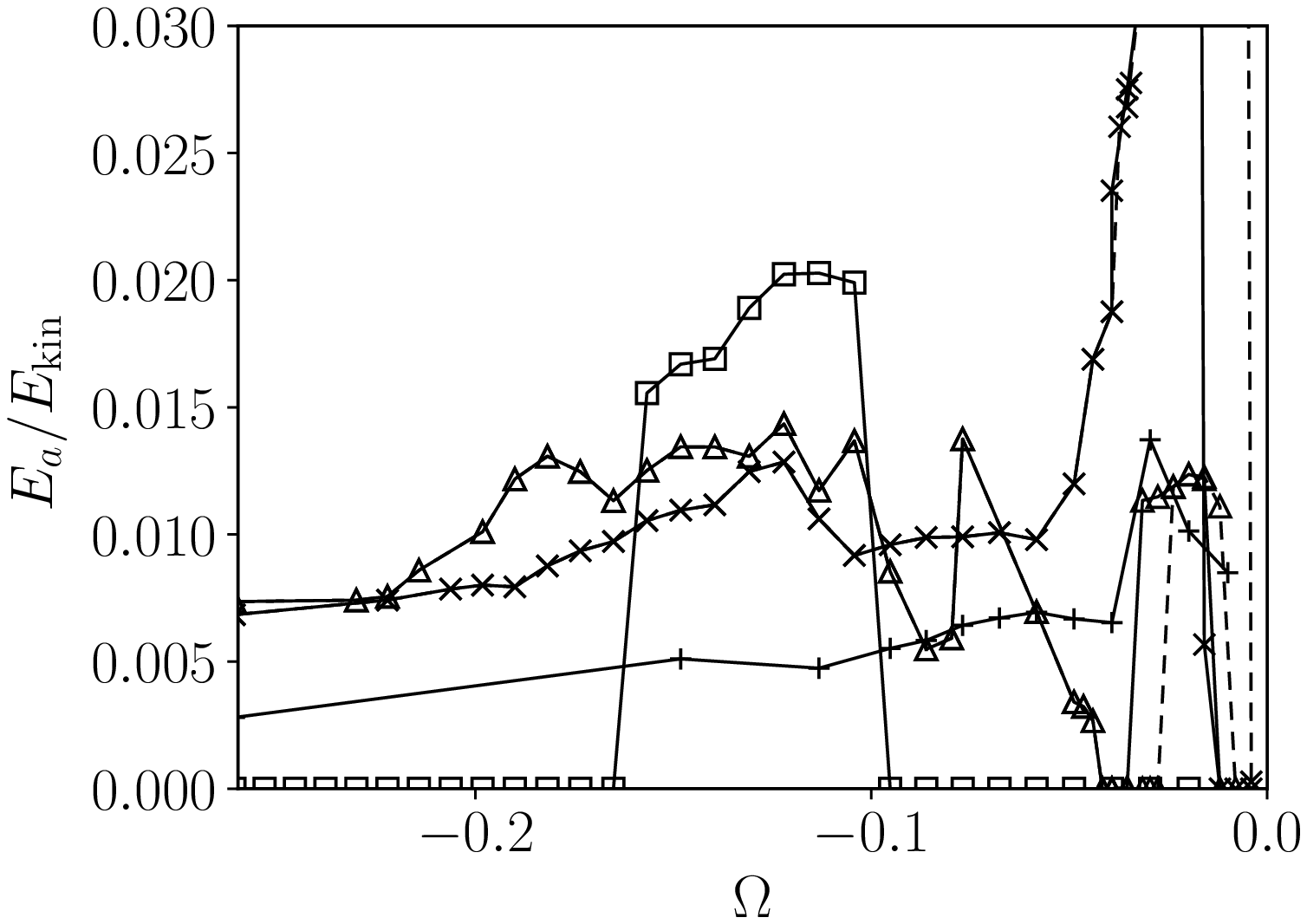}}}%
\caption{$E_a/E_{\mathrm{kin}}$ as a function of $\Omega$ for
$\mathrm{Ek}=1.0\cdot10^{-3}$ ($\square$), $\mathrm{Ek}=2.5\cdot10^{-4}$ ($\triangle$),
$\mathrm{Ek}=1.0\cdot10^{-4}$ ($\times$) and $\mathrm{Ek}=1.0\cdot10^{-5}$ ($+$).
The dashed line is followed when decreasing $|\Omega|$, the continuous line is
obtained in going in the opposite direction. The right panel shows the same data as
the left panel on a different scale.}%
\label{fig5}
\end{minipage}
\end{center}
\end{figure}

The antisymmetric components $\bm v_a$ can be excited only via an instability. Their
energy $E_a$, defined in (\ref{eq:E_a}), is therefore a convenient 
indicator for the presence of instability.
Figure~\ref{fig5} shows $E_a/E_{\mathrm{kin}}$ for comparison with
figures~\ref{fig2} and \ref{fig3}. It is seen that for $\mathrm{Ek}=10^{-3}$, the
interval of $\Omega$ in which $E_a/E_{\mathrm{kin}}\neq 0$ coincides with the
interval in which significant axisymmetric components are present in the
meridional flow. In fact, at
this $\mathrm{Ek}$, there is no other instability than the one leading to the
$s_2t_1$ and $s_1t_1$ flows. At the other $\mathrm{Ek}$, however, the flow first
becomes unstable through triad resonances \citep{Goepfe16} and the $s_2t_1$ and
$s_1t_1$ flows exist side by side with inertial modes.

\section{Kinematic dynamos}
\label{Kinematic}

There are several possible definitions of the magnetic Reynolds number which are
potentially of interest and which differ in the velocity on which they are
based. The definition which is most directly related to the parameters of an
experiment is the magnetic Reynolds number computed from the rotational velocity
of the container about its axis, $\mathrm{Rm}_\mathrm{rot}$, given by
\begin{equation}
\mathrm{Rm}_\mathrm{rot}\, = \,\mathrm{Pm}/(2 \mathrm{Ek})\,.
\end{equation}
Structural stability and the available motors naturally set a limit on the largest
$\mathrm{Rm}_\mathrm{rot}$ achievable in an experiment, which happens to be 1420
in the Dresden experiment \citep{Stefan12,Stefan15}.

\begin{figure}
\begin{center}
\includegraphics[width=100mm]{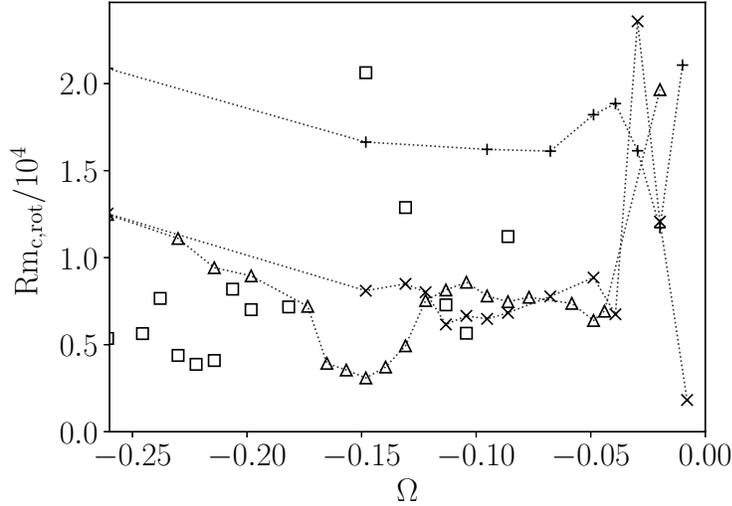}
\caption{$\mathrm{Rm}_\mathrm{c,rot}$ as a function of $\Omega$ for
$\mathrm{Ek}=1.0\cdot10^{-3}$ ($\square$), $\mathrm{Ek}=2.5\cdot10^{-4}$ ($\triangle$),
$\mathrm{Ek}=1.0\cdot10^{-4}$ ($\times$) and $\mathrm{Ek}=1.0\cdot10^{-5}$ ($+$).}
\label{fig6}
\end{center}
\end{figure}

The critical value of this magnetic Reynolds number,
$\mathrm{Rm}_\mathrm{c,rot}$, is shown for the various simulations in figure~\ref{fig6}.
For all $\mathrm{Ek} < 10^{-3}$, a triad resonance occurs. Inertial modes
in triads are known to be able to generate magnetic fields, so that these triads are
responsible for a baseline in this figure and also for some of the salient
variations. For instance, the best dynamo in figure~\ref{fig6} is realized at 
$\mathrm{Ek} = 10^{-4}$ and $\Omega=-0.008$, which is within a hysteresis loop so
that this flow must be accessed by lowering $|\Omega|$ from higher values (see
figure~\ref{fig5}). The $\mathrm{Rm}_\mathrm{c,rot}$ is then 1820. At these parameters,
the energy in the antisymmetric and hence unstable modes is exceptionally large as
can be seen in figure~\ref{fig5}, whereas the axisymmetric energy stays small
according to figures~\ref{fig2} and \ref{fig3}, so that this dynamo is driven by a triad.

There are other notable variations in $\mathrm{Rm}_\mathrm{c,rot}$ in figure~\ref{fig6}
which correlate with axisymmetric flow components.
Figures~\ref{fig2}(a) and \ref{fig3}(a)
tell us whether an axisymmetric flow of large amplitude comes on top
of the inertial modes, and we can
deduce from figure~\ref{fig3}(b) whether this flow is mostly of the $s_2t_1$ structure
or whether there are large contributions by the $s_1t_1$ flow. The recognizable
dips in the curve representing $\mathrm{Rm}_\mathrm{c,rot}$ in figure~\ref{fig6}
occur in intervals of $\Omega$ in which 
an $s_2t_1$ flow of significant amplitude is present (for example for
$\mathrm{Ek} = 2.5 \times 10^{-4}$ around $\Omega=-0.15$). 
If on the contrary there is a large contribution by a flow of
the $s_1t_1$ type (at $\mathrm{Ek} = 10^{-3}$ around
$\Omega=-0.15$), the dynamo worsens. The axisymmetric flows thus have an effect
on magnetic field generation, but they are not necessarily helpful. In the available
examples, the $s_2t_1$ component tends to lower $\mathrm{Rm}_\mathrm{c,rot}$
whereas $s_1t_1$ acts in the opposite direction.
 
\begin{figure}
\begin{center}
\includegraphics[width=100mm]{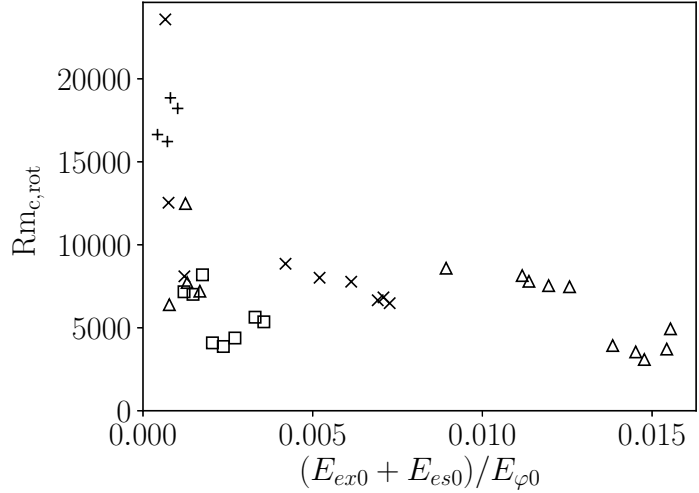}
\caption{$\mathrm{Rm}_\mathrm{c,rot}$ as a function of
$(E_{x0}+E_{s0})/E_{\varphi 0}$ for dynamos in flows with
$(E_{ex0}+E_{es0}) > \frac{1}{2} (E_{x0}+E_{s0})$, which selects flows
dominated by the $s_2t_1$ flow in their axisymmetric components.
The different markers stand for
$\mathrm{Ek}=1.0\cdot10^{-3}$ ($\square$), $\mathrm{Ek}=2.5\cdot10^{-4}$ ($\triangle$),
$\mathrm{Ek}=1.0\cdot10^{-4}$ ($\times$) and $\mathrm{Ek}=1.0\cdot10^{-5}$ ($+$).}
\label{fig7}
\end{center}
\end{figure}

It is known from optimization studies done in connection with the VKS experiment
\citep{Ravele05} that axisymmetric flows of the type studied by \citet{Dudley89}
are most effective at generating magnetic fields if their poloidal and toroidal
energies are comparable. The flows studied here all have 
$(E_{x0}+E_{s0})/E_{\varphi 0} < 0.016$ (see figure~\ref{fig3}(a)) and must be
inefficient according to this criterion. Figure~\ref{fig7} collects all the
dynamos with a significant $s_2t_1$ flow. The critical
magnetic Reynolds number is on the order of several thousands as opposed to
one hundred for the optimized flows in \citet{Ravele05}, and the
critical magnetic Reynolds number decreases with increasing 
$(E_{x0}+E_{s0})/E_{\varphi 0}$ at small $(E_{x0}+E_{s0})/E_{\varphi 0}$.

While the $s_2t_1$ flow helps dynamo action, its presence does not lower the
critical magnetic Reynolds number in our examples to a value accessible in the Dresden
experiment.

\begin{figure}
\begin{center}
\includegraphics[width=100mm]{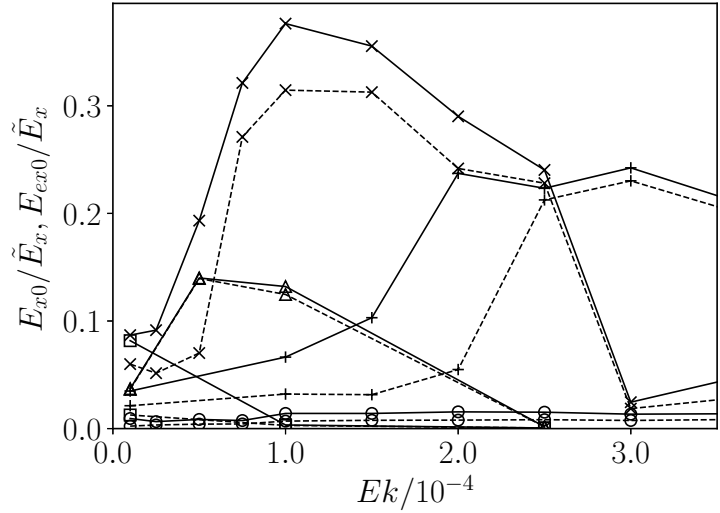}
\caption{$E_{x0}/\tilde E_x$ as a function of $\mathrm{Ek}$ for 
$\Omega=-0.1$ ($\times$), $\Omega=-0.15$ ($+$) and $\Omega=-0.26$ ($\circ$).
The continuous lines connect points for $E_{x0}/\tilde E_x$, and the dashed lines are
for $E_{ex0}/\tilde E_x$.}
\label{fig8}
\end{center}
\end{figure}

\begin{figure}
\begin{center}
\includegraphics[width=100mm]{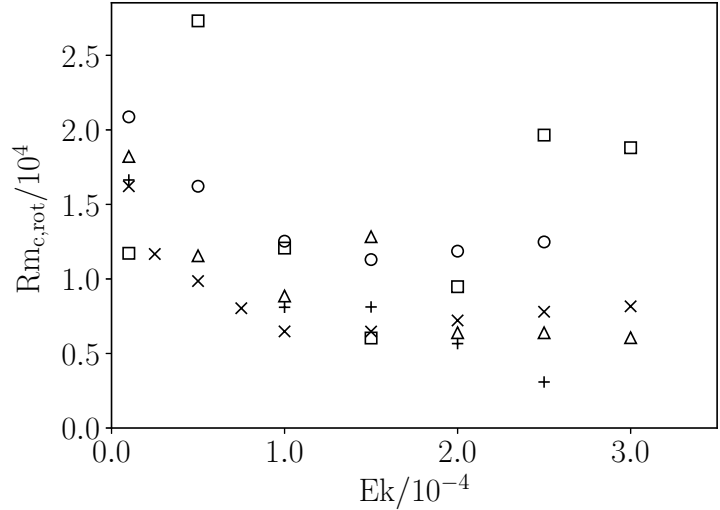}
\caption{$\mathrm{Rm}_\mathrm{c,rot}$ as a function of $\mathrm{Ek}$ for 
$\Omega=-0.02$ ($\square$), $\Omega=-0.05$ ($\triangle$), $\Omega=-0.1$ ($\times$),
$\Omega=-0.15$ ($+$) and $\Omega=-0.26$ ($\circ$).
}
\label{fig9}
\end{center}
\end{figure}

Simulations of both astrophysical objects and experiments generally have the
problem that they cannot simulate the small Ekman numbers which are of interest.
The Dresden experiment for instance can be operated at $\mathrm{Ek}$ as low as
$10^{-8}$, whereas all our simulations are at $\mathrm{Ek} \geq 10^{-5}$. The behavior
at small $\mathrm{Ek}$ has to be deduced from extrapolations. Ideally, the
extrapolation is based on theory and physical understanding. In order to safely
extrapolate the dynamo properties of the $s_2t_1$ flow, we would need to know
how it is excited, and whether it will persist at small $\mathrm{Ek}$. The
mechanism exciting this flow is not elucidated. The pattern of the 
$s_2t_1$ flow is compatible with a centrifugal instability, as proposed
by \citet{Giesec18}. Whatever the
true mechanism may be, it seems to become inoperative at low $\mathrm{Ek}$. As
figure~\ref{fig8} shows, $E_{x0}/\tilde E_x \rightarrow 0$ as $\mathrm{Ek} \rightarrow 0$ at
any fixed $\Omega$, so that we have to expect the beneficial effects of the
$s_2t_1$ flow for the dynamo to disappear at small $\mathrm{Ek}$. This is part
of the reason why critical magnetic Reynolds numbers generally increase with
decreasing $\mathrm{Ek}$, as shown in figure~\ref{fig9}. However, also the dynamos
among our simulations
operating with triadic resonances deteriorate with decreasing $\mathrm{Ek}$, so
that there must be yet another reason for this behavior.

\begin{figure}
\begin{center}
\includegraphics[width=100mm]{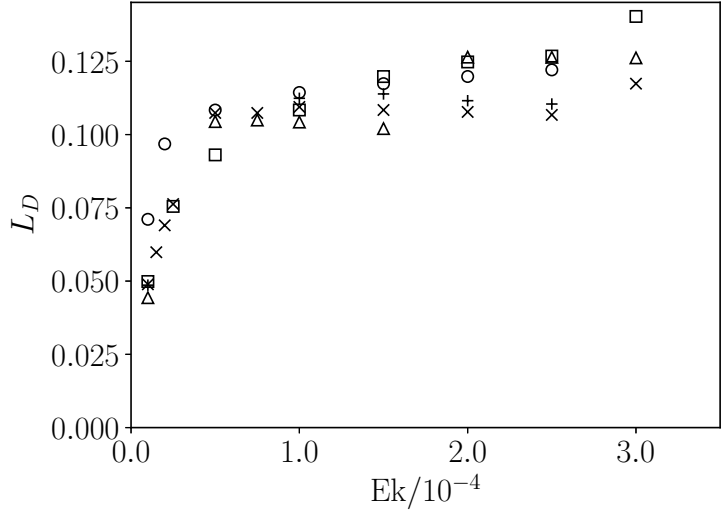}
\caption{$L_D$ as a function of $\mathrm{Ek}$ for 
$\Omega=-0.02$ ($\square$), $\Omega=-0.05$ ($\triangle$), $\Omega=-0.1$ ($\times$),
$\Omega=-0.15$ ($+$) and $\Omega=-0.26$ ($\circ$).}
\label{fig10}
\end{center}
\end{figure}

\begin{figure}
\begin{center}
\includegraphics[width=100mm]{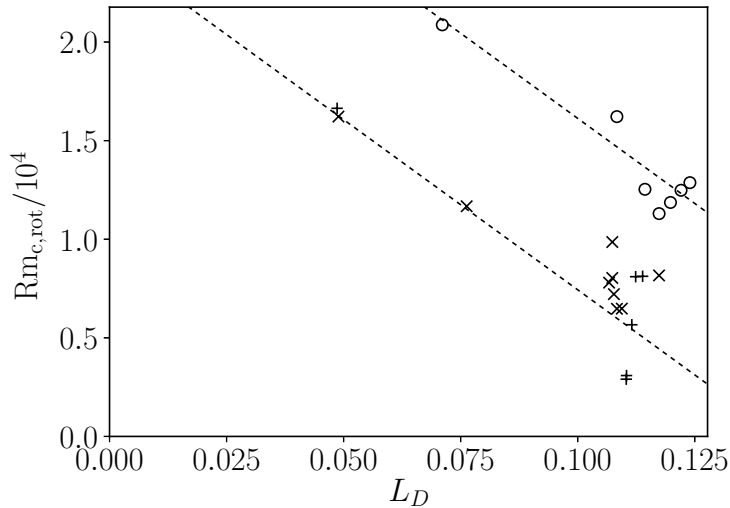}
\caption{$\mathrm{Rm}_\mathrm{c,rot}$ as a function of $L_D$.
The different markers stand for $\Omega=-0.1$ ($\times$),
$\Omega=-0.15$ ($+$) and $\Omega=-0.26$ ($\circ$). The upper dashed lines guide the eye
through the points for $\Omega=-0.26$ and the other through all remaining points.}
\label{fig11}
\end{center}
\end{figure}

Another contribution to this effect may come from increasing turbulence and the
appearance of small scales at small $\mathrm{Ek}$. Let us use a
dissipation length scale $L_D$ as diagnostics for the presence of small scale
structures. The dissipation $D$ is given by
\begin{equation}
D\, =\, \mathrm{Ek}\, \frac{1}{V} \biggl\langle\, \int \sum_{ij} (\upartial_i v_j)^2 \, {\mathrm d}V  \biggr\rangle
\,=\, \frac{1}{V} \, \biggl\langle \,\int \bm v {\bm \cdot} \bigl( \Omega_P \omega_D (\hat{\bm x}
\times \bm p(t)) \times \bm r \bigr)\, {\mathrm d}V \biggr\rangle\,.
\end{equation}
The last equation results from taking the scalar product of (\ref{eq:NS_BQ})
with $\bm v$, integrating over space and averaging over time. While both
expressions for $D$ are identical analytically, the first expression incurs the
larger numerical error because it depends on derivatives, so that the second
expression was always used to extract $D$ from the numerical results.
Finally, the dissipation length is defined as
\begin{equation}
L_D \,=\, \sqrt{E_\mathrm{kin}/D}\,.
\end{equation}
As expected, $L_D$ is approximately constant in the laminar flows and decreases
with decreasing $\mathrm{Ek}$ at small $\mathrm{Ek}$ (see figure~\ref{fig10}). The
decrease in $L_D$ correlates with the increase of $\mathrm{Rm}_\mathrm{c,rot}$
in figure~\ref{fig9}. To make this clear, figure~\ref{fig11} plots
$\mathrm{Rm}_\mathrm{c,rot}$ directly as function of $L_D$. Studies in dependence
of $\mathrm{Ek}$ in precessing cubes are complicated because the flow may
undergo transitions between different states as $\mathrm{Ek}$ is varied
\citep{Goepfe16}, for example between different triad resonances, or a triad
resonance and a single vortex state, or a significant axisymmetric component may
come and go. This is particularly true of the points with $\Omega=-0.02$ 
which are therefore not shown in figure~\ref{fig11}. At least at the
precession rates included in figure~\ref{fig11}, $\mathrm{Rm}_\mathrm{c,rot}$ varies as a function of $L_D$ for
fixed $\Omega$ on lines parallel to each other, suggesting that the eddy
diffusivity introduced by turbulence is increasing the critical magnetic
Reynolds number.

\section{Conclusion}
\label{Conclusion}

Several mechanisms enabling precession driven flows to act as dynamos have been
identified in the past. Ekman pumping at the boundaries and triad resonances
were the first to be observed \citep{Tilgne05}. Dynamos in long slender vortices
which form at low Ekman numbers were found later \citep{Goepfe16}. Recently
\citep{Giesec18}, it was shown that for certain parameters, the flow in
precessing cylinders resembles the $s_2t_1$ flow studied by \citet{Dudley89} as
kinematic dynamo in a sphere. The present work confirms the appearance of this
$s_2t_1$ flow to be a common feature in precessing flows. The $s_2t_1$ flow
helps in generating magnetic fields, although not to the extent that the results from
precessing cubes allow one to propose parameters at which the Dresden
experiment should act as a dynamo. We also find $s_1t_1$ flows in the cube.
There is no theory yet as to what drives these flows. It is therefore not
possible to safely extrapolate the numerical results to small Ekman numbers. A purely
empirical extrapolation is difficult because the flow transits between different
states as the Ekman number is lowered at fixed precession rate. Generally,
however, the critical magnetic Reynolds number increases in our sinulations
with decreasing Ekman
number at low Ekman numbers. This increase occurs in parallel to the appearance
of turbulence and small scales in the flow. In addition, the strong axisymmetric
flow components disappear at small Ekman numbers.


\end{document}